\documentclass[12pt]{article}
\usepackage{epsfig}
\usepackage{amsmath}
\usepackage{hhline}
\usepackage{amssymb}
\usepackage{times}
\usepackage{cite}
\usepackage{lineno}

\newlength{\dinwidth}
\newlength{\dinmargin}
\begin{document}  
\newcommand{\gev}{\ensuremath{\mathrm{GeV}}}
\newcommand{\gevsq}{\ensuremath{\mathrm{GeV}^2}}
\newcommand{\mus}{\ensuremath{\mu}s}

\newcommand{\sqrts}{$\sqrt{s}$}
\newcommand{\shat}{\hat{s}}
\newcommand{\pt}{\ensuremath{P_{T}}}
\newcommand{\yjb}{\ensuremath{y_{JB}}}  
\newcommand{\qsq}{\ensuremath{Q^2}}
\newcommand{\qzerosq}{\ensuremath{Q^2_0}}
\newcommand{\alphas}{\ensuremath{\alpha_s}}

\newcommand{\dglap}{\ensuremath{\alpha_s^n \ln^n Q^2}}
\newcommand{\bfkl}{\ensuremath{\alpha_s^n \ln^n (1/x)}}

\newcommand{\deltaeta}{\ensuremath{\Delta\eta}}
\newcommand{\pom}{{I\!P}}
\newcommand{\reg}{{I\!R}}
\newcommand{\modt}{\ensuremath{|t|}}
\newcommand{\massx}{\ensuremath{M_X}}
\newcommand{\massy}{\ensuremath{M_Y}}
\newcommand{\PO}{I\!\!P} 
\newcommand{\RO}{I\!\!R}
\newcommand{\xpom}{\ensuremath{x_{\PO}} }

\newcommand{\cts}{\ensuremath{\cos \theta^{\ast}}}
\newcommand{\ts}{\ensuremath{\theta^{\ast}}}
\newcommand{\ps}{\ensuremath{\phi^{\ast}}}
\newcommand{\rfour}{\ensuremath{r^{04}_{00}}}
\newcommand{\rone}{\ensuremath{r^{04}_{10}}}
\newcommand{\rreal}{\ensuremath{{\rm Re}\left [r^{04}_{10} \right ]}}
\newcommand{\rmone}{\ensuremath{r^{04}_{1-1}}}
\newcommand{\app}{\ensuremath{M_{++}}}
\newcommand{\apo}{\ensuremath{M_{+0}}}
\newcommand{\apm}{\ensuremath{M_{+-}}}
\newcommand{\mpipi}{\ensuremath{M_{\pi\pi}}}
\newcommand{\mkk}{\ensuremath{M_{KK}}}
\newcommand{\mrho}{\ensuremath{m_{\rh}}}
\newcommand{\wrho}{\ensuremath{\Gamma_{\rh}}}
\newcommand{\abfkl}{\ensuremath{\alpha_s^{BFKL}}}
\newcommand{\aif}{\ensuremath{\alpha_s^{IF}}}
\newcommand{\pzrh}{\ensuremath{P_{z, \rh}}}
\newcommand{\ptrh}{\ensuremath{P_{t, \rh}}}
\newcommand{\erh}{\ensuremath{E_{\rh}}}
\newcommand{\trange}{\ensuremath{|t| > 1.5~\gev}}
\newcommand{\myrange}{\ensuremath{M_{Y} < 5~\gev}}
\newcommand{\wrange}{\ensuremath{75 < W < 95~\gev}}
\newcommand{\qrange}{\ensuremath{\qsq < 0.01~\gevsq}}
\newcommand{\powerlaw}{\ensuremath{n~=~4.26~\pm~0.06\stat~^{+0.06}_{-0.04}\syst}}
\newcommand{\zeuspowerlaw}{\ensuremath{n~=~3.21~\pm~0.04\stat~\pm 0.15\syst}}

\newcommand{\jpsi}{\ensuremath{J/\psi}}
\newcommand{\rh}{\ensuremath{\rho}}
\newcommand{\rhdecay}{\ensuremath{\rho \rightarrow \pi^+ \pi^-}}
\newcommand{\om}{\ensuremath{\omega}}
\newcommand{\ph}{\ensuremath{\phi}}
\newcommand{\rhprime}{\ensuremath{\rho^{\prime}}}

\newcommand{\gapprox}{\raisebox{-0.7ex}{$\stackrel {\textstyle>}{\sim}$}}
\newcommand{\lapprox}{\raisebox{-0.7ex}{$\stackrel {\textstyle<}{\sim}$}}
\def\gsim{\,\lower.25ex\hbox{$\scriptstyle\sim$}\kern-1.30ex%
\raise 0.55ex\hbox{$\scriptstyle >$}\,}
\def\lsim{\,\lower.25ex\hbox{$\scriptstyle\sim$}\kern-1.30ex%
\raise 0.55ex\hbox{$\scriptstyle <$}\,}
\def\lapproxeq{\lower .7ex\hbox{$\;\stackrel{\textstyle<}{\sim}\;$}}
\def\gapproxeq{\lower .7ex\hbox{$\;\stackrel{\textstyle>}{\sim}\;$}}
\newcommand{\gap}{\stackrel{>}{\sim}}
\newcommand{\lap}{\stackrel{<}{\sim}}

\newcommand{\gp}{\ensuremath{\gamma p}}
\newcommand{\ep}{\ensuremath{ep}}
\newcommand{\stat}{\ensuremath{~\rm (stat.)}}
\newcommand{\syst}{\ensuremath{~\rm (syst.)}}
\newcommand{\gavin}{Poludniowski}

\def\Journal#1#2#3#4{{#1} {\bf #2} (#3) #4}
\def\NCA{\em Nuovo Cimento}
\def\NIM{\em Nucl. Instrum. Methods}
\def\NIMA{{\em Nucl. Instrum. Methods} {\bf A}}
\def\NPB{{\em Nucl. Phys.}   {\bf B}}
\def\PLB{{\em Phys. Lett.}   {\bf B}}
\def\PRL{\em Phys. Rev. Lett.}
\def\PRD{{\em Phys. Rev.}    {\bf D}}
\def\ZPC{{\em Z. Phys.}      {\bf C}}
\def\EJC{{\em Eur. Phys. J.} {\bf C}}
\def\CPC{\em Comp. Phys. Commun.}

\begin{titlepage}

\noindent
\begin{flushleft}
DESY 06-023\hfill ISSN 0418-9833\\
March 2006
\end{flushleft}

\vspace{2cm}

\begin{center}
\begin{Large}

{\bf  Diffractive Photoproduction of \boldmath{\rh}\ Mesons \linebreak 
with Large Momentum Transfer at HERA}

\vspace{2cm}

H1 Collaboration

\end{Large}
\end{center}

\vspace{2cm}

\begin{abstract}
\noindent
The diffractive photoproduction of \rh\ mesons, $\ep \rightarrow e\rh Y$, with large momentum transfer squared  
at the proton vertex, \modt, is studied with the H1 detector at HERA using an integrated luminosity 
of $20.1 \ \rm pb^{-1}$.  The photon-proton centre of mass energy spans the range \wrange , the photon 
virtuality is restricted to \qrange\ and the mass $M_Y$ of the proton remnant is below 5 GeV.  
The $t$ dependence of the cross section is measured for the range $1.5 < \modt < 10.0$~\gevsq\ and 
is well described by a power law, ${\rm d}\sigma/{\rm d}\modt \propto \modt^{-n}$.  The spin density matrix 
elements, which provide information on the helicity structure of the interaction, are extracted using 
measurements of angular distributions of the $\rho$ decay products.  The data indicate a violation of $s$-channel helicity 
conservation, with contributions from both single and double helicity-flip being observed.  The results are
compared to the predictions of perturbative QCD models.
\end{abstract}

\vspace{1.5cm}

\begin{center}
Submitted to \PLB
\end{center}

\end{titlepage}

%
%
%
\begin{flushleft}

A.~Aktas$^{9}$,                
V.~Andreev$^{25}$,             
T.~Anthonis$^{3}$,             
B.~Antunovic$^{26}$,           
S.~Aplin$^{9}$,                
A.~Asmone$^{33}$,              
A.~Astvatsatourov$^{3}$,       
A.~Babaev$^{24, \dagger}$,     
S.~Backovic$^{30}$,            
A.~Baghdasaryan$^{37}$
% ...

\bigskip{\it
% ...
$ ^{3}$ Inter-University Institute for High Energies ULB-VUB, Brussels;
          Universiteit Antwerpen, Antwerpen; Belgium$^{ c}$ \\
% ...
$ ^{9}$ DESY, Hamburg, Germany \\
% ...
$ ^{24}$ Institute for Theoretical and Experimental Physics,
          Moscow, Russia$^{ k}$ \\
 $ ^{25}$ Lebedev Physical Institute, Moscow, Russia$^{ e}$ \\
 $ ^{26}$ Max-Planck-Institut f\"ur Physik, M\"unchen, Germany \\
% ...
$ ^{30}$ Faculty of Science, University of Montenegro,
          Podgorica, Serbia and Montenegro$^{ e}$ \\
% ...
$ ^{33}$ Dipartimento di Fisica Universit\`a di Roma Tre
          and INFN Roma~3, Roma, Italy \\
% ...
$ ^{37}$ Yerevan Physics Institute, Yerevan, Armenia \\
% ...
$ ^{\dagger}$ Deceased \\

\bigskip
% ...
$ ^c$ Supported by FNRS-FWO-Vlaanderen, IISN-IIKW and IWT
      and  by Interuniversity
Attraction Poles Programme,
      Belgian Science Policy \\
% ...
$ ^e$ Supported by the Deutsche Forschungsgemeinschaft \\
% ...
$ ^k$ Partially Supported by Russian Foundation
      for Basic Research,  grants  03-02-17291
      and  04-02-16445 \\
}
\end{flushleft}

\newpage

\section{Introduction}

Diffractive vector meson production in \ep\ interactions with large negative four-momentum transfer squared at the proton 
vertex, $t$, provides a powerful means to probe the nature of the diffractive exchange.
It has been proposed as a process in which Quantum Chromodynamics (QCD) effects predicted by the BFKL  evolution
equation~\cite{forshawryskin,bartels} could be observed.
Theoretically, the large momentum transfer provides the hard scale necessary for the 
application of perturbative QCD (pQCD) models.  In such models, diffractive vector meson production is viewed, in the proton 
rest frame, as a sequence of three processes well separated in time: the intermediate photon fluctuates into a $q\bar{q}$ 
pair; the $q\bar{q}$ pair is involved in a hard interaction with the proton via the exchange of a colour singlet 
state, and the $q\bar{q}$ pair recombines to form a vector meson.  At leading order (LO), the interaction 
between the $q\bar{q}$ pair and the proton is represented by the exchange of two gluons.  Beyond LO, the exchange of 
a gluon ladder has to be considered which, in the leading logarithm (LL) approximation, can be described by the 
BFKL equation. The treatment of the formation of the vector meson, which is a non-perturbative process, involves a parameterisation 
of the meson wavefunction.

At high \modt , the $t$ dependence of vector meson production and of the spin density matrix elements has been measured in 
photoproduction by ZEUS~\cite{zeushight} (\rh, \ph, \jpsi) and H1~\cite{hightjpsi,hightpsi} (\jpsi, $\psi (2s)$). 
H1 has also measured the $t$ dependence 
of the spin density matrix elements for the \rh\ meson at high \modt\ in electroproduction~\cite{electrosdme}.
The data on light vector meson production (\rh, \ph) at large \modt\ indicate a violation of $s$-channel 
helicity conservation (SCHC), i.e.\ the
non-conservation of the helicity between the exchanged photon and the vector meson.  This is in contrast 
to measurements of the heavier \jpsi\ meson, where $s$-channel helicity is seen to be conserved~\cite{zeushight,hightjpsi}.

This paper presents new measurements of the diffractive production of $\rh$ mesons at large \modt, in the range 
$1.5 < \modt < 10\ {\rm GeV^{2}}$:
\begin{equation}
\ep \rightarrow e\rh Y; \rh \rightarrow \pi^+\pi^-,
\label{eq:reaction} 
\end{equation} 
in the photoproduction regime, i.e.\ $\qsq \simeq 0$~\gevsq\ where \qsq\ is the modulus of the squared 
four momentum carried by the intermediate photon.  The system $Y$ represents either an elastically 
scattered proton or a low mass dissociated system.   A clean signature and low background rates make 
this process experimentally attractive and it is possible to precisely
determine the kinematics of the process through an accurate measurement of the vector meson four
momentum. 
The dependence on  $t$ of the cross section is measured 
and the spin density matrix elements are extracted, which provide information on the helicity 
structure of the interaction.

\section{Perturbative QCD Models} \label{sec:theory}

Perturbative QCD calculations at leading order (two-gluon exchange) have
been performed for high \modt\ vector meson photoproduction in~\cite{Serbo}.
For light vector mesons, these calculations indicate that, although the
initial photon is transversely polarised, the vector meson is produced
predominantly in a longitudinal state. Intuitively, this can be understood
as follows. Since the light quark and antiquark have opposite helicities
(chiral-even configuration), the pair must be in an orbital momentum
state with projection $L_z = \pm 1$ onto the photon axis, in order to conserve
the photon spin projection. The hard interaction between the dipole and
the gluon system does not affect the dipole size nor the quark and
antiquark helicities, but modifies the dipole line of flight. This implies
a damping by a factor $\propto 1 / \modt$ of the probability of measuring the value
$L_z = \pm 1$ for the projection of the dipole angular momentum onto its line
of flight, and hence of transversely polarised vector meson production.

However, in contrast to these pQCD expectations, experimental observations indicate 
that light vector mesons are produced predominantly in a transverse polarisation state [3].
As first realised in~\cite{ivanov} an enhanced production of transversely polarised 
$\rho$ mesons can be accounted for 
by the possibility for a real photon to couple to a $q \bar{q}$ pair in
a chiral-odd spin configuration even in the case of
        light quarks\footnote{
                Note that in the case of heavy vector mesons, the relevant
                non-relativistic wavefunction, with equal sharing of the photon longitudinal momentum 
		by the two heavy quarks, ensures that the amplitude for
                producing a longitudinally polarised meson vanishes.
                Only the chiral-odd amplitude, which is naturally
                present due to the non-zero quark mass, remains. Hence heavy vector mesons
                are expected to be transversely polarised, as
                confirmed by the experimental observations.}.

The data presented in this paper are compared to two theoretical predictions: 
a fixed order calculation in which the hard interaction is approximated by the exchange of two 
gluons\footnote{The two-gluon model predictions for the kinematic region considered here were provided by the
        authors of~\cite{jeff1,jeff2}.  The model differs from that 
        proposed in~\cite{ivanov} in the way the chiral-odd configurations are introduced.},
and a LL calculation in which it is described according to the BFKL evolution.
In both cases, the chiral-odd component
of the photon wavefunction is obtained by giving the quarks a ``constituent quark mass"
$m = m_V/2$, where $m_V$ is the mass of the vector meson, and
a set of QCD light-cone wavefunctions for the vector meson~\cite{Ball:1998ff} is used.

The BFKL calculation is described in~\cite{jeff1,jeff2}. The BFKL resummation cures some 
possible instabilities which might affect the two-gluon predictions~\cite{forshawryskin}.
The expansions on the light-cone of the relevant hadronic matrix elements are performed up
to twist-3, i.e.\ next-to-leading twist.
The leading logarithm nature of the calculation prevents an absolute prediction for the normalisation of the cross
sections, due to the presence of an undefined energy scale $\Lambda$.  In~\cite{jeff1,jeff2} this scale is allowed
to run with $t$ according to $\Lambda^2 = m_V^2 - \gamma t$, where $\gamma$ is a free parameter.  The value of the 
strong coupling constant is fixed since this is appropriate in the LL approximation and has proved successful~\cite{jeffnonrel,cox} 
in describing previous data.  The parameter values, as obtained from a fit to the ZEUS data in~\cite{zeushight}, are:
$\gamma = 1.0$ and $\abfkl = 0.20$~\cite{enberg}.

\section{Data Analysis} \label{sec:ana}
\subsection{Event Selection} \label{sec:sel}

The data used for the present analysis were taken with the H1 detector in the year 2000 and correspond 
to an integrated luminosity of $20.1~{\rm pb^{-1}}$. In this period, the energies of the HERA proton 
and positron beams were 920 GeV and 27.5 GeV, respectively. The kinematic domain of the measurement is:
\begin{eqnarray}
&\qsq& < 0.01\ {\rm GeV^2}     \nonumber \\
75 < &W& < 95\ {\rm GeV}       \nonumber \\
1.5 < &\modt& < 10\ {\rm GeV^{2}}   \nonumber \\
&M_Y & < 5\ {\rm GeV}      
\label{eq:kin} 
\end{eqnarray} 
where 
$W$ is the photon-proton centre of mass energy and $M_Y$ is the mass of the 
proton remnant system.  

The relevant parts of the detector, for which more details can be found in~\cite{nim}, are the central 
tracking detector (which covers the polar angular range $20^{\circ} < \theta < 160^{\circ}$), the liquid 
argon (LAr) calorimeter (which covers the polar angular range $4^{\circ} < \theta < 154^{\circ}$) and an 
electron tagger located at $44~\rm{m}$ from the interaction point, which detects the scattered positron at a small 
angle to the backward direction\footnote{In the H1 convention, the $z$ axis is defined by the 
colliding beams, the forward direction being that of the outgoing proton beam and the backward 
direction that of the positron beam.  Transverse and longitudinal momenta are defined with respect 
to the proton beam direction.}.  The $44~\rm m$ electron tagger is a $2 \times 3$ array of \v{C}erenkov 
crystal calorimeters used to select photoproduction events in the range $\qsq < 0.01$~\gevsq.  The trigger 
system used in this analysis selects events with an energy deposit  greater than $10$~GeV in the $44~\rm m$ electron 
tagger, at least one charged track with a transverse momentum above $400$~MeV in the central tracker and 
a reconstructed interaction vertex. Additionally, there is a veto on the amount of energy deposited in the 
forward region of the LAr.

Events corresponding to reaction~(\ref{eq:reaction}), in the kinematic range defined by relations~(\ref{eq:kin}), 
are selected by requesting:
\begin{itemize}
\item
the reconstruction of an energy deposit of more than 15~\gev\ in the $44~\rm m$ electron tagger (the 
scattered positron candidate);

\item
the reconstruction in the central tracking detector of the trajectories of exactly two 
oppositely charged particles (pion candidates) with transverse momenta larger than $150$~MeV 
and polar angles within  $20^{\rm o} < \theta < 155^{\rm o}$.  In order to 
ensure a well understood trigger efficiency, at least one track is required to have a transverse 
momentum above $450$~MeV;

\item
the absence of any localised energy deposit larger than 400~MeV in the LAr calorimeter 
which is not associated with either of the two reconstructed tracks.  This cut reduces backgrounds due 
to the diffractive production of systems decaying into two charged and additional neutral particles.
Further, it limits the mass of the proton dissociative system to $M_Y~\lapprox~5$ GeV which ensures 
that the events lie within the diffractive regime $M_Y \ll W$. Restricting the analysis to low values
of $M_Y$ also reduces the uncertainties arising from the parametrisation of the $M_Y$ dependence of the 
cross section;

\item
events in the mass range $0.6 < \mpipi < 1.1$~\gev\ and discarding events with \linebreak$\mkk < 1.04$~\gev, 
where \mpipi\ and \mkk\ are the invariant masses of the two selected tracks when considered 
as pions or kaons respectively (no explicit hadron identification is performed for this analysis). The 
latter cut reduces the background due to diffractive production of \ph\ mesons.
\end{itemize}

The final sample consists of 2628 events.  Further details of this analysis may be found in~\cite{carl}.

\subsection{Kinematics and Helicity Structure} \label{sec:kinematics}

The three momentum of the \rh\ meson is computed as the sum of the two charged pion candidate momenta.
The variable $W$ is reconstructed from the \rh\ meson, rather than from the energy of the scattered 
positron candidate which is less precisely measured by the electron tagger, using the Jacquet-Blondel method~\cite{jb}:
\begin{equation}
W^2 \simeq 2E_{p}(E_{\rho} - p_{z, \rho})
\label{eq:w}
\end{equation}
where $E_p$ is the energy of the incoming proton and $E_{\rh}$ and $p_{z,\rh}$ are the energy 
and longitudinal momentum of the \rh\ meson, respectively. In the photoproduction regime, the 
variable $t$ is well approximated by the negative transverse momentum squared of the \rh\ meson, 
$t \simeq -p_{t,\rh}^2$.

\begin{figure}[f]
  \centering
  \setlength{\unitlength}{1cm}
  {\epsfig{file=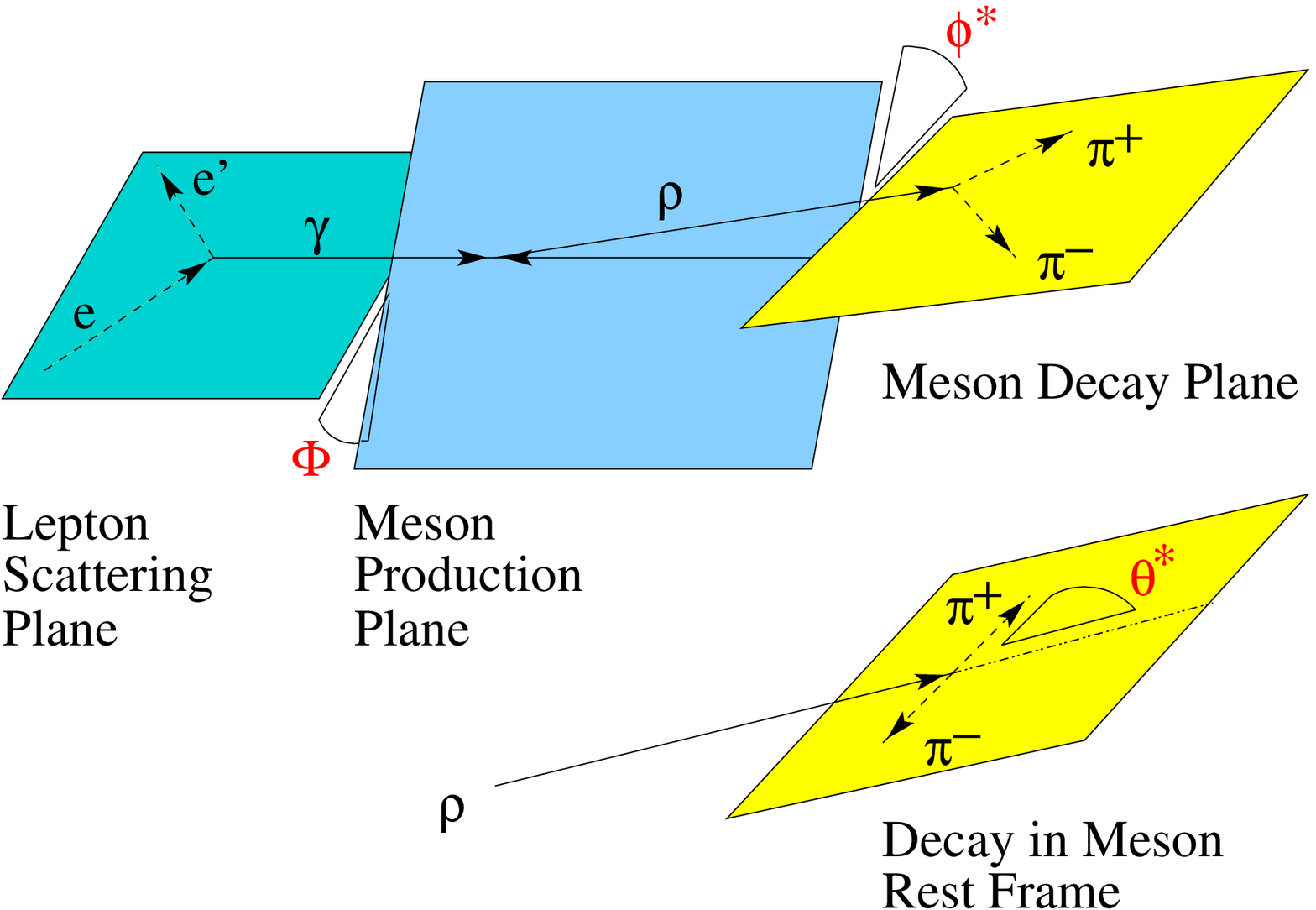, width=12.0cm}}
  \caption{Production and decay angles used to analyse the polarisation of 
	the \rh\ meson.}
  \label{fig:plane}
\end{figure}

The measurement of the production and decay angular distributions provides information 
on the helicity structure of the interaction.  Three angles are defined~\cite{angles} as illustrated in 
Fig.~\ref{fig:plane}: $\Phi$ is 
the angle between the \rh\ production plane (defined as the plane containing the virtual photon and the 
\rh\ meson) and the positron scattering plane in the \gp\ centre of mass system; \ts\ is the polar 
angle of the positively charged decay pion in the \rh\ rest frame with respect to the meson direction as defined in 
the \gp\ centre of mass frame and \ps\ is its azimuthal angle relative to the \rh\ production plane.

In this paper, the distributions of the angles \ps\ and \ts\ are analysed (the angle $\Phi$ is not 
accessible in photoproduction) giving access to the spin density matrix elements \rfour, \rreal\ and \rmone~\cite{sdme}.
For a \rh\ meson decaying into two pions, the normalised two-dimensional angular distribution~\cite{sdme}, 
averaged over $\Phi$, is 
\begin{eqnarray}
\frac{1}{\sigma} \frac{\rm{d}^2\sigma}{\rm{d} \cos\theta^{\ast}d\phi^{\ast}} &=& \frac{3}{4\pi} \left [
\frac{1}{2} (1 - \rfour) + \frac{1}{2} (3\rfour - 1) \cos^2\theta^{\ast} \right . \nonumber \\
&&\left . -\sqrt{2}\rreal \sin2\theta^{\ast}\cos\phi^{\ast} - \rmone\sin^2\theta^{\ast}\cos 2\phi^{\ast}
\right ].
\label{eqn:helicity}
\end{eqnarray}
Integrating over \cts\ or \ps\  further reduces this distribution to the one dimensional distributions 
\begin{equation}
\frac{{\rm d}\sigma}{{\rm d}\cos \theta^{\ast}} \propto 1 - \rfour + (3\rfour - 1)\cos^2\theta^{\ast}
\label{eq:ctseqn}
\end{equation}
and
\begin{equation}
\frac{{\rm d}\sigma}{{\rm d}\phi^{\ast}} \propto 1 - 2\rmone \cos 2\phi^{\ast}.
\label{eq:pseqn}
\end{equation}
The spin density matrix 
elements are defined as bilinear combinations of the helicity amplitudes $M_{\lambda_{\gamma}\lambda_{V}}$, 
where $\lambda_{\gamma},\lambda_{V} = -,0,+$ are the respective helicities of the photon and the vector meson.  
For photoproduction, where the photon is quasi-real, the longitudinal photon polarisation component 
is negligible and only the transverse polarisation states remain.  The photon-meson transitions can thus be 
described in terms of three independent helicity amplitudes \app, \apo, \apm\footnote{
The three corresponding amplitudes $M_{--}$, $M_{-0}$ and $M_{-+}$ are not independent 
since they satisfy $\app = M_{--}$, $\apm = M_{-+}$ and $\apo = -M_{-0}$ due to parity symmetry.}, which correspond to no  
change in helicity (no-flip), a single change in helicity (single-flip) and a double change in helicity 
(double-flip), respectively.  In this case, the matrix elements are related to the helicity amplitudes 
by
\begin{eqnarray}
\rfour &=& \frac{|M_{+0}|^2}
{|M_{++}|^2 + |M_{+0}|^2 + |M_{+-}|^2} \label{eqn:r0400} \\
\rone &=& \frac{1}{2} \frac{M_{++}M_{+0}^{\ast} - M_{+-}M_{+0}^{\ast}}
{|M_{++}|^2 + |M_{+0}|^2 + |M_{+-}|^2} \label{eqn:r0410} \\
\rmone &=& \frac{1}{2} \frac{M_{++}M_{+-}^{\ast} + M_{+-}M_{++}^{\ast}}
{|M_{++}|^2 + |M_{+0}|^2 + |M_{+-}|^2}. \label{eqn:r0411}
\label{eqn:helamps}
\end{eqnarray}
Under $s$-channel helicity conservation, only the amplitude \app\ is non-zero 
and consequently the \rfour, \rone\ and \rmone\ matrix elements should all be zero.
In contrast, both the BFKL and two-gluon models predict a 
violation of SCHC.  In the case of the LL BFKL model, the helicity amplitudes are 
predicted to follow a hierarchical structure with $|\app| > |\apm| > |\apo|$~\cite{jeff1,jeff2}.

\subsection{ Monte Carlo Simulation} \label{sec:MC}

A Monte Carlo simulation based on the DIFFVM program~\cite{diffvm} is used to describe   
the diffractive production and decay of $\rho$ mesons, and to correct the data 
for acceptance, efficiency and smearing effects.  The \qsq\ and $W$ dependences of the cross section 
are taken from previous measurements~\cite{diffvmparam}. The $t$ dependence is 
taken according to the power law measured in the present analysis following an iterative procedure. 
The simulation includes the angular distributions corresponding to the measurements of the present 
analysis for the \rfour, \rmone\ and the \rreal\ matrix elements.   
The $M_Y$ spectrum is parameterised as 
${\rm d} \sigma / {\rm d} M_Y^2 \propto f(M_Y^2)/M_Y^{2.15}$~\cite{goulianos},
where $f(M_Y^2) = 1$ for $M_Y^2 > 3.6$ \gevsq\ and, at lower masses, is  a function  
which accounts for the production of excited nucleon states.  Other angular distributions 
and correlations are taken in the {\it s}-channel helicity conservation approximation.
The mass distribution is described by a relativistic Breit-Wigner distribution, 
the mass and the width of the $\rho$ being fixed~\cite{pdg}, including 
skewing effects resulting from the interference with open pion pair production \cite{soding,rs} 
taken from the current analysis. For studies of systematics uncertainties, all simulation 
parameters have been varied within errors (see section~\ref{sec:syst}).

The DIFFVM simulation is also used for the description of the \om, \ph\ and \rhprime\ 
backgrounds (see next section).  Here the $t$ distributions are described using the same 
parameterisation as for the \rh\ meson and the angular 
distributions are kept in the SCHC approximation, which is justified by the smallness
of these contributions.

\begin{figure}[f]
\begin{center}
\setlength{\unitlength}{1cm}
\epsfig{file=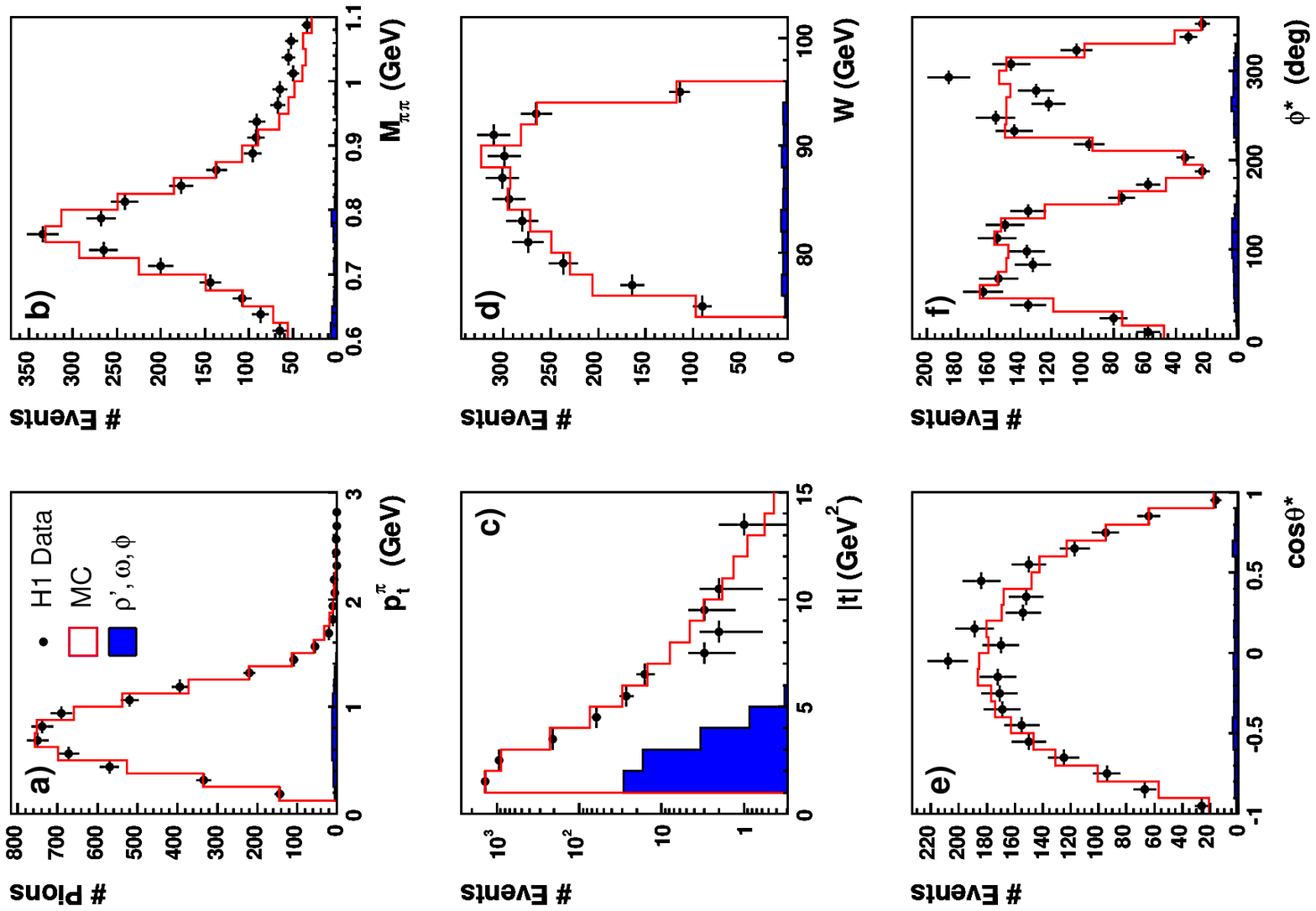,angle=270,width=13.8cm}\\ 
\end{center}
\vspace{-0.6cm}
 \caption{Distributions of the selected events in the kinematic domain (\protect\ref{eq:kin}) and 
the invariant mass range $0.6 < \mpipi < 1.1$~\gev:
a) transverse momentum of the pions; b) invariant mass of the two pions;
c) modulus of the square of the four momentum transferred at the proton vertex $t$; d) photon-proton 
centre of mass energy $W$; e) cosine of the polar angle \ts\  of the positively charged decay pion in the
$\rho$ rest frame and f)  its azimuthal angle \ps\  
 the \gp\ centre of mass frame.  The points represent the data and the 
histograms show the Monte-Carlo predictions normalised to the data, including the \om , \ph\ and \protect\rhprime\ 
backgrounds (filled histograms).}
 \label{fig:control}
 \end{figure}

All generated samples are passed through a detailed simulation of the detector
response based on the GEANT3 program~\cite{geant}, and through the same 
reconstruction software as used for the data. 
Figure~\ref{fig:control} presents the observed and simulated distributions for several 
variables of the selected sample of events.  The simulated distributions, which  include
the amounts of background discussed in section~\ref{sec:bkg}, are normalised to the number of observed events. 
Reasonable agreement is observed for all distributions, 
showing that the Monte-Carlo simulations can reliably be used to correct the data 
for acceptance and smearing effects.  The structure in the \ps\ distribution (Fig.~\ref{fig:control}f) 
is due to the low geometrical acceptance of the central tracking
detector in the case where the angle between the \rh\ meson production and decay 
planes is small ($\ps \sim 0^{\circ}$ or $180^{\circ}$), which leads to the
emission of one of the pions at a small angle relative to the beam direction.

\subsection{Backgrounds} \label{sec:bkg}

Diffractive photoproduction of \om, \ph\ and \rhprime\ mesons\footnote{The detailed structure~\cite{pdg} of 
the \rhprime\ state is not relevant for the present study. The 
name \rhprime\ is, therefore, used to represent both the \rhprime(1450) and the \rhprime(1700). In the DIFFVM 
simulation, the \rhprime\ mass and width are taken as $1450$~MeV and $300$~MeV, respectively.} can fake 
\rh\ production through the decay channels:
\begin{eqnarray}
\om \rightarrow \pi^+ \pi^- \pi^0,&&\nonumber\\
\ph \rightarrow \pi^+ \pi^- \pi^0, &\ph \rightarrow K^0_S K^0_L,& \nonumber \\ 
\rho^\prime \rightarrow \rho^{\pm} \pi^{\mp} \pi^0, &\rh^{\pm} \rightarrow \pi^{\pm} \pi^0,&
\label{eq:bgs}
\end{eqnarray}
if the decay photons of the $\pi^0$ or the $K^0_L$ mesons are not detected\footnote{
The contribution from background processes leading to more than two charged particles in
the final state is negligible.}. This happens in the 
cases where the energy of the neutral particle is deposited in an inactive region of the detector, is associated 
to the charged pion tracks, or is below the noise threshold.  Diffractive photoproduction of \om\ 
and $\phi$ mesons also gives the same topology as that of the \rh\ meson within the detector through the decay 
channels
\begin{eqnarray}
\om &\rightarrow& \pi^+\pi^-,\nonumber\\
\ph &\rightarrow& K^+K^-.
\end{eqnarray}

To estimate the corresponding backgrounds, 
the \om, \ph\ and \rhprime\ cross sections were taken from measured ratios to the \rh\ cross section 
in the \qsq\ range relevant for the analysis: $\sigma_{\om}~/~\sigma_{\rh} = 0.106 \pm 0.019$~\cite{omega}, 
$\sigma_{\ph}~/~\sigma_{\rh} = 0.156 _{-0.019}^{+0.029}$~\cite{zeushight} and 
$\sigma(\rhprime  \rightarrow \rh^{\pm} \pi^{\mp} \pi^0)~/~\sigma_{\rh} = 0.2 \pm 0.1$. In 
the latter case, the ratio is obtained from the ratio 
$\sigma(\rhprime \rightarrow \rho^0 \pi^+ \pi^-)~/~\sigma_{\rh}
= 0.10 \pm 0.05$, measured in electron~\cite{prime1} and muon~\cite{prime2} scattering off a liquid hydrogen target, 
under the assumption
\begin{equation}
\frac{\sigma(\rhprime \rightarrow \rho^+ \pi^- \pi^0) + 
\sigma(\rhprime \rightarrow \rho^- \pi^+ \pi^0)}
{\sigma(\rhprime \rightarrow \rho^0 \pi^+ \pi^-)}  = 2.
\end{equation}

The background contributions in the selected kinematic domain~(\ref{eq:kin}) and for the selected \rh\ 
mass range are estimated and subtracted separately for each measurement interval using the Monte Carlo 
simulations.  In total, the backgrounds amount to  0.5\%, 0.2\% and 1.2\% for \om, \ph\ and 
\rhprime\ production, respectively. 

\subsection{Systematic Uncertainties} \label{sec:syst}

The systematic uncertainties on the measurements are estimated by varying the event selection, the 
parameters of the \rh\ Monte Carlo simulation and the properties of subtracted
backgrounds. 
The following sources of systematic error are taken 
into account:

\begin{itemize}
\item {\bf Uncertainties in the \boldmath{\rh}\ simulation} \\
The uncertainty on the input $t$ distribution, used to compute acceptances and smearing effects and to adjust the 
measurements to the mean $t$ value for each $t$ interval, is taken into account by varying the exponent of the 
Monte Carlo power law by $\pm 0.5$.  The uncertainty in the modelling of the dissociative 
proton system $Y$ is estimated by reweighting the proton remnant mass distribution by factors 
$(1/M_Y^2)^{\pm 0.3}$~\cite{goulianos}.  For the angular distributions, the spin density matrix elements are varied around 
the values measured in the current data according to the spread of the observed results with $t$ by $\pm~0.03$ 
for \rfour, $\pm~0.02$ for \rreal\ and $^{+0.02}_{-0.04}$ for \rmone. Finally, in the low \modt\ region where 
non-zero skewing is observed in the \rh\ line shape, the skewing parameter is varied according to the uncertainty 
of the fit to the invariant mass distribution (see section~\ref{sec:results}).

\item {\bf Uncertainties on the background distributions} \\
The amount of background is varied by changing the ratio of the background cross sections to the \rh\ 
cross section according to their uncertainties, as quoted in section~\ref{sec:bkg}.  
The $t$ dependence of the \rhprime\ distribution, which provides the largest background contribution, 
is further varied using weighting factors of $(1/\modt)^{\pm 2.0}$.

\item {\bf Uncertainties in the detector description} \\
Uncertainties on the detailed $p_t$ and angular dependences of the 
trigger efficiencies are taken into account by varying them within their estimated errors.  
The uncertainty on the tracking acceptance at large angles is estimated by varying 
the cut on the polar angle of the reconstructed tracks between $150^\circ$ and $160^\circ$.  
The threshold for the detection of energy deposits in the LAr that are not associated to the two pion candidates is 
varied between $300$~MeV and $500$~MeV.  Finally, the influence of the uncertainty in the description of 
the electron tagger acceptance is estimated by shifting the acceptance range in $W$ by 3\%.

\end{itemize}

For the measurement of the $t$ dependence, the largest sources of systematic uncertainty are the slope of the $t$ distribution 
in the MC, the variation of the LAr energy threshold and the variation of the upper $\theta$ cut.  Additionally, for the measurement
of the spin density matrix elements, the parameterisation of the matrix elements in the MC provides a significant 
effect.  For each contribution, the relative effect on the extracted $t$ slope is less than $\pm 1\%$, and the effect 
on the measured spin density matrix elements is less than $\pm 0.015$.

The total systematic error on the cross section is obtained by adding the individual contributions, which are
considered as uncorrelated, in quadrature.  Correlated systematics which affect only the normalisation cancel 
since only normalised cross sections are presented here. For the extraction of the $t$ slope and the spin density 
matrix elements, the systematic error is obtained by repeating the appropriate fit after shifting the data points 
according to each individual systematic uncertainty. Again, the individual contributions are added in quadrature.

\section{Results} \label{sec:results}

In each bin of the kinematic variables, the cross section is computed from the number of events 
in the bin, fully corrected for backgrounds, acceptance and smearing effects using the Monte Carlo 
simulations described above.  At low \modt , the data are further corrected for the skewing effect obtained 
from a fit to the present data using the Ross-Stodolsky~\cite{rs} parameterisation in which the mass
and the width of the $\rho$ are fixed~\cite{pdg}. This correction amounts to $2.6 \%$ for $ 1.5 < \modt < 2.2 \ \gevsq $. 
At higher \modt, the skewing effect is negligible. All cross sections presented 
below, as well as the theoretical model predictions, are normalised to their integrals in the respective 
kinematic domain.

\subsection{Dependence on \boldmath{$t$}} \label{sec:t}

\begin{figure}[f]
\vspace{-0cm}
\begin{center}
\epsfig{file=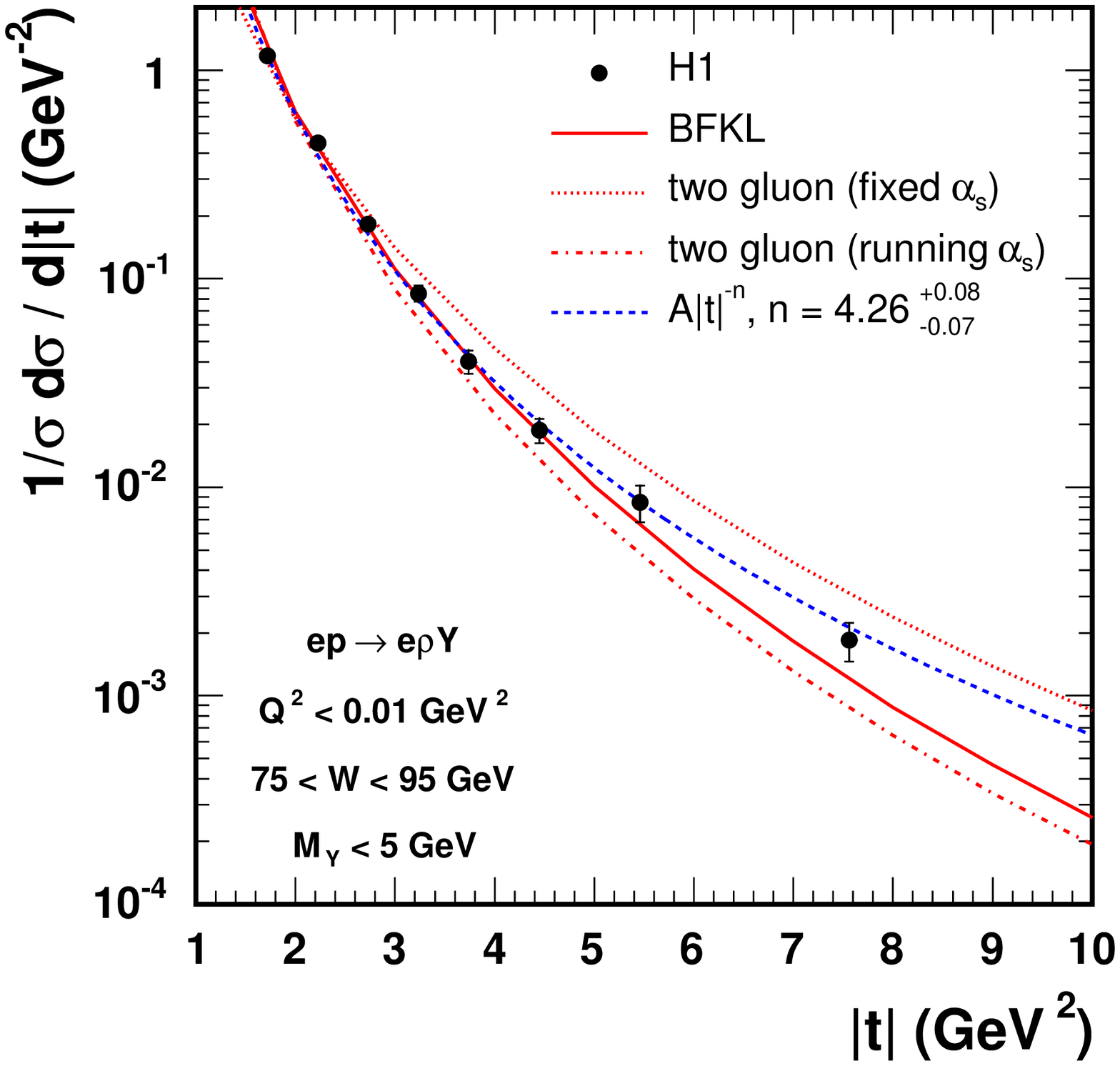,angle=270,width=16.cm}\\ 
\end{center}
 \caption{The $t$ dependence of the $\ep \rightarrow e\rh Y$ cross section.  The inner error bars show 
the dominating statistical errors, while the outer error bars represent the sum of the statistical and  
systematic errors added in quadrature.  The dashed line is the result of a fit to a power law 
distribution $|t|^{-n}$, which results in a power \powerlaw.  The full line shows the prediction 
from the BFKL model and the dotted and dashed-dotted lines show the predictions from the two-gluon model 
with fixed and running $\alpha_s$, respectively. } 
 \label{fig:t}
 \end{figure}

The $t$ dependence of the $\ep \rightarrow e\rh Y$ cross section is presented in Fig.~\ref{fig:t} and table~\ref{tab:t}.  
The data are plotted at the mean value in each $t$ interval determined according to the parameterisation of the 
$t$ dependence.  The data are well described over the measured range of $t$ by a power law dependence of the form 
${\rm d}\sigma/{\rm d}\modt \propto |t|^{-n}$ where \powerlaw, as determined by a $\chi^2$ minimisation.  An exponential 
parameterisation of the form ${\rm d}\sigma/{\rm d}\modt \propto e^{-b\modt}$ is unable to describe the data over the full $t$ 
range. The data in Fig.~\ref{fig:t} are compared with the predictions of the two-gluon model both with fixed and 
running $\alpha_s$ and with those of the BFKL model~\cite{jeff1,jeff2}. The BFKL model provides a reasonable description 
of the $t$ dependence, in contrast to the two-gluon model predictions.

The ZEUS Collaboration has previously published data on the diffractive photoproduction of \rh\ mesons with 
proton dissociation in the range $1.1 < |t| < 10.0$~\gevsq\ \cite{zeushight} and observes a power law with 
exponent \zeuspowerlaw.  This is significantly shallower than the result obtained here.  The difference 
in slope can be understood in terms of the difference in the kinematic region over which the two measurements 
are made, in particular the maximum value of $M_Y$: the phase space of the ZEUS measurement corresponds to
 $M_Y < 12$~\gev\ ($31$~\gev) at $|t| = 1.5$~\gevsq\ ($|t| = 10$~\gevsq) whereas the phase space of this measurement
is $M_Y < 5$ \gev. 
Both measurements are well described by the BFKL model using similar parameters~\cite{jeff2,enberg}.

\begin{table}[tbp]
\centering
\begin{tabular}{|c|c|c|c|c|}
\hline
&&\\[-6pt]
\modt\ range&$\langle \modt \rangle$&$1/\sigma~{\rm d}\sigma/{\rm d}\modt$\\
(\gevsq)&(\gevsq)&($\rm GeV^{-2}$)\\[6pt]
\hline\hline
&&\\[-6pt]
$1.5 - 2.0$&$1.72$&$1.176 \pm 0.032 ~_{-0.012} ^{+0.020}$\\[6pt]
$2.0 - 2.5$&$2.23$&$0.447 \pm 0.019 ~_{-0.018} ^{+0.0051}$\\[6pt]
$2.5 - 3.0$&$2.73$&$0.183 \pm 0.012 ~_{-0.0062} ^{+0.0072}$\\[6pt]
$3.0 - 3.5$&$3.23$&$0.0850 \pm 0.0076 ~_{-0.0041} ^{+0.0025}$\\[6pt]
$3.5 - 4.0$&$3.74$&$0.0401 \pm 0.0051 ~_{-0.0019} ^{+0.0031}$\\[6pt]
$4.0 - 5.0$&$4.45$&$0.0188 \pm 0.0025 ~_{-0.0009} ^{+0.0014}$\\[6pt]
$5.0 - 6.0$&$5.46$&$0.00848 \pm 0.00167 ~_{-0.00034} ^{+0.00066}$\\[6pt]
$6.0 - 10.0$&$7.56$&$0.00185 \pm 0.00039 ~_{-0.00009} ^{+0.00011}$\\[6pt]
\hline
\end{tabular}
\caption{The normalised differential cross section for $\ep \rightarrow e\rh Y$ as a function of $t$. The first errors are 
	statistical and the second are systematic. The kinematic range of the measurement is given in~(\ref{eq:kin}).}
\label{tab:t}
\end{table}

\subsection{Spin Density Matrix Elements} \label{sec:sdme}

\begin{figure}[f]
\vspace{-0.cm}
\begin{center}
\setlength{\unitlength}{1cm}
\epsfig{file=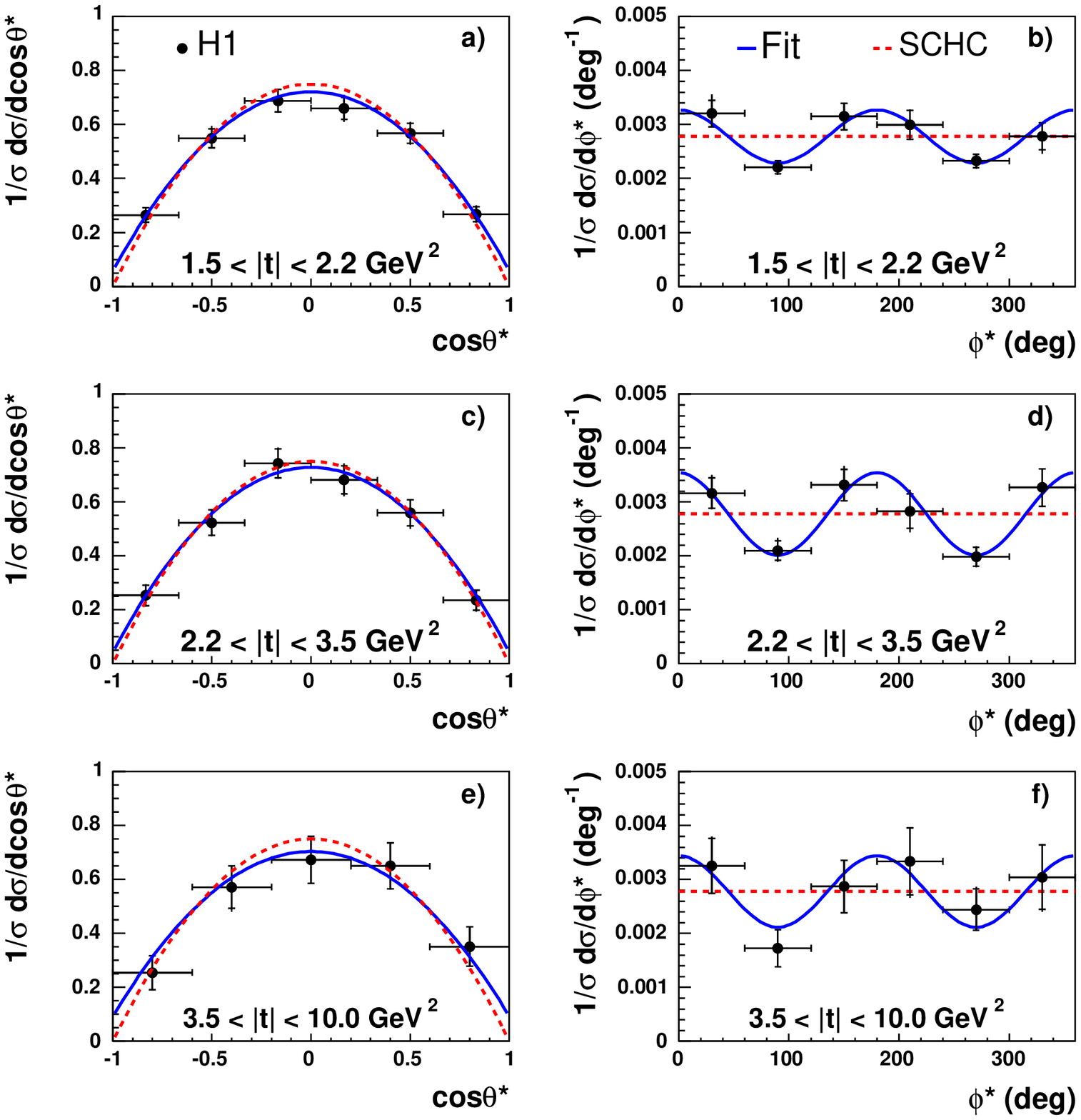,angle=270,width=16.cm}\\ 
\end{center}
 \caption{Normalised decay angular distributions for \rh\ meson photoproduction in three bins of \modt.  
The left column (a,c,e) shows the polar distribution \cts\ and the right column (b,d,f) shows 
the azimuthal distribution \ps .  The inner error bars show the statistical errors, 
while the outer error bars represent the sum of the statistical and systematic errors 
added in quadrature.  The solid lines show the results of the two-dimensional fit to the data (see 
text).  The dashed lines show the expectations for $s$-channel helicity conservation (SCHC).}
 \label{fig:angles}
 \end{figure}

\begin{figure}[f]
\vspace{-0.cm}
\begin{center}
\setlength{\unitlength}{1cm}
\epsfig{file=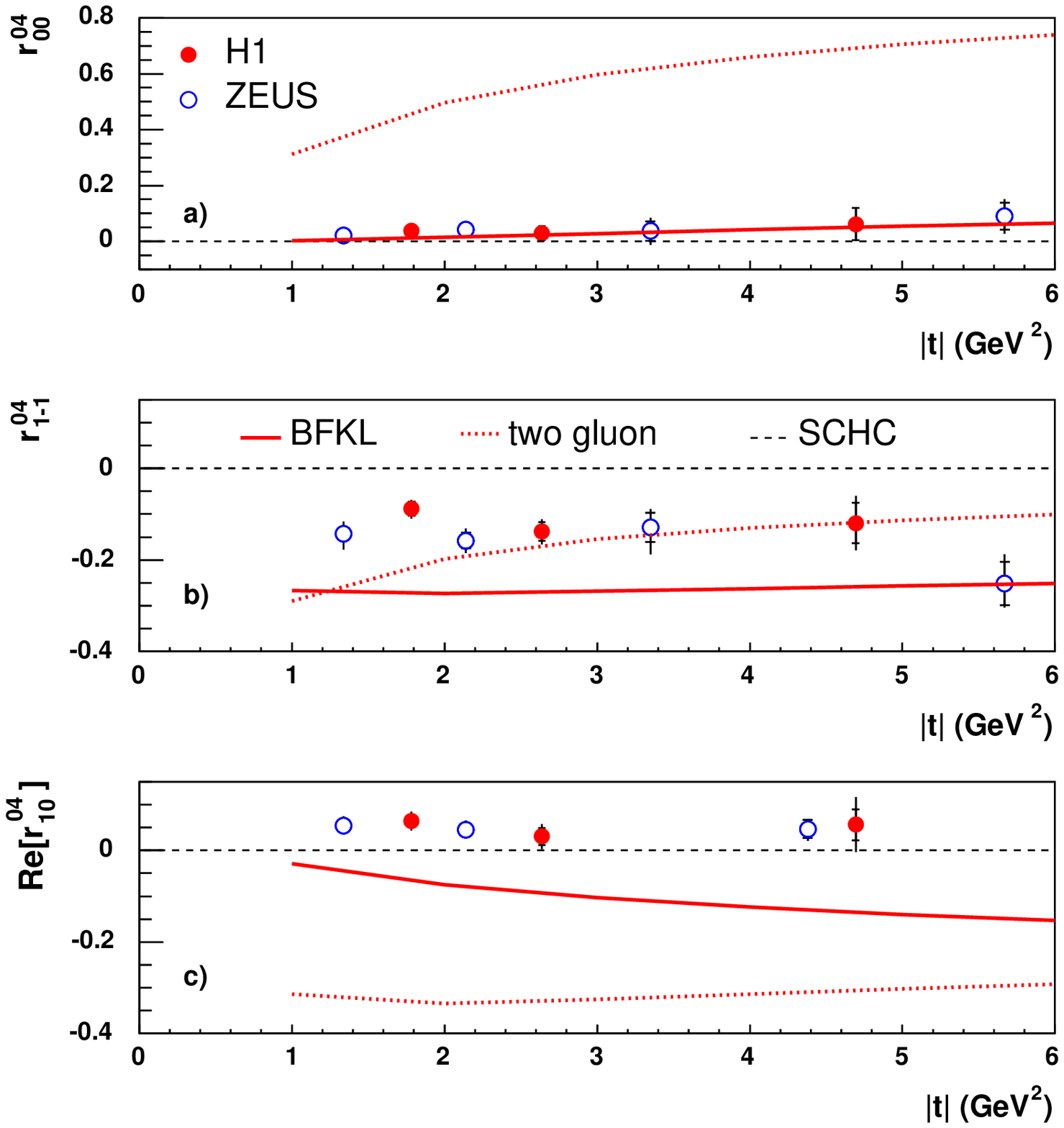,angle=270,width=16.cm}\\ 
\end{center}
 \caption{The three spin density matrix elements a) \rfour, b) \rmone\ and c) \rreal\ for  
\rh\ meson photoproduction as a function of \modt\ (full points) together with ZEUS measurements 
\cite{zeushight} (open points). The inner error bars show the statistical errors, while the 
outer error bars represent the sum of the statistical and systematic errors added in quadrature. 
The full lines show the predictions of the BFKL model and the dotted lines show the predictions 
of the two-gluon model.  The dashed lines show the expectation from $s$--channel 
helicity conservation (SCHC).}
 \label{fig:sdme}
 \end{figure}

The spin density matrix elements are extracted by a two-dimensional likelihood fit of equation~(\ref{eqn:helicity})
to the data.  The normalised single differential distributions in \cts\ and \ps\ 
are shown in Fig.~\ref{fig:angles} for three ranges of $t$.  The solid curves show the projection 
of the two-dimensional fit and the dashed curves show the expectation of $s$-channel helicity 
conservation.  A flat \ps\ behaviour is clearly disfavoured, indicating a violation of SCHC. The 
values of the three extracted matrix elements are shown in Fig.~\ref{fig:sdme} and table~\ref{tab:sdme} 
as a function of \modt.  Within the experimental uncertainty, no strong dependence on $t$ is observed.
Measurements of the spin density matrix elements for the photoproduction of \rh\ 
mesons obtained by the ZEUS Collaboration~\cite{zeushight} are also shown.  There is a reasonable agreement between 
the results of the two experiments. 

The small values of \rfour, which is directly proportional to the square of the single-flip helicity amplitude 
\apo, signify that the probability of producing a longitudinally polarised \rh\ from a transversely polarised photon 
is low, varying from $(4 \pm 2)\%$ at $\modt = 1.79$~\gevsq\ to $(6 \pm 6)\%$ at $\modt = 4.69$~\gevsq.  The non-zero 
values of \rreal\ confirm that, although small, a single-flip contribution is present.  The production of 
transversely polarised \rh\ mesons must, therefore, dominate and the finite negative values of \rmone\ show clear 
evidence for a helicity double-flip contribution.  Both these observations indicate a violation of the SCHC 
hypothesis.  This is in contrast with the results on the \jpsi\ meson, where, within experimental errors, the measured 
spin density matrix elements~\cite{hightjpsi,zeushight} are all compatible with zero. 

The two-gluon model predictions (dotted lines in Fig.~\ref{fig:sdme})~\cite{jeff1,jeff2} 
are unable to describe the measured spin density matrix elements. In particular, the model predicts
too high values of \rfour , i.e. too high probabilities for producing longitudinally polarised \rh\ mesons.
For the BFKL predictions~\cite{jeff1,jeff2}, which 
are shown by the full lines in Fig.~\ref{fig:sdme}, \rfour\ is well described but the prediction for 
\rmone\ is too negative and the wrong sign for \rreal\ is predicted. The inability to describe the \rreal\ 
matrix element is the major obstacle for the BFKL model.


\section{Summary} \label{sec:conc}

The diffractive  photoproduction of $\rho$ mesons, $\ep \rightarrow e\rho Y$, has been studied 
using the H1 detector at HERA in the kinematic range $Q^2 < 0.01 $~\gevsq , $75~<~W~<~95$~GeV, 
$1.5~<~|t|~<~10$~\gevsq\ and $M_Y~<~5~\rm GeV$. The $t$ dependence of the cross section is 
measured and fitted with a power law of the form $|t|^{-n}$, which fits the data well and results in \powerlaw .
It is reasonably described by a BFKL-based model, while both two-gluon predictions considered here, 
with different treatments of the strong coupling, fail to describe the data.

The spin density matrix elements \rfour, \rmone\ and \rreal\ are measured as a function of $t$. 
The \rmone\ and \rreal\ matrix elements differ significantly from zero, thus confirming the violation 
of $s$-channel helicity conservation, with contributions from both single and double helicity-flip observed. 
The models considered are unable to describe the spin density matrix elements.  The two-gluon model predicts 
far too large a probability of producing a longitudinally polarised \rh\ meson, given by the \rfour\ matrix 
element.  While the BFKL based model is able to describe the \rfour\ matrix element well, the prediction for 
\rmone\ is too negative and \rreal\ has the wrong sign.


\begin{table}[tbp]
\centering
\begin{tabular}{|c|c|c|c|c|}
\hline
&          &      &                                    &\\[-6pt]
\modt\ range&$\langle \modt \rangle$&\rfour\ & \rreal\ &\rmone\\
(\gevsq)&(\gevsq)&&&\\[6pt]
\hline\hline
&          &      &                                    &\\[-6pt]
$1.5 - 2.2$&$1.79$&$0.038 \pm 0.017~_{-0.012}^{+0.011}$&$0.064 \pm
0.012~_{-0.015}^{+0.005}$&$-0.088 \pm 0.015~_{-0.014}^{+0.007}$\\[6pt]
$2.2 - 3.5$&$2.64$&$0.029 \pm 0.025~_{-0.013}^{+0.010}$&$0.031 \pm
0.019~_{-0.011}^{+0.007}$&$-0.138 \pm 0.021~_{-0.011}^{+0.011}$\\[6pt]
$3.5 - 10.0$&$4.69$&$0.062 \pm 0.058~_{-0.012}^{+0.015}$&$0.057 \pm
0.034~_{-0.007}^{+0.004}$&$-0.119 \pm 0.044~_{-0.009}^{+0.011}$\\[-6pt]
&          &      &                                    &\\
\hline
\end{tabular}
\caption{
The three spin density matrix elements \rfour, \rmone\ and \rreal\ for \rh\ 
meson photoproduction as a function of \modt. The first errors are statistical 
and the second are systematic.  }
 \label{tab:sdme}
\end{table}

\section*{Acknowledgements}

We are grateful to the HERA machine group whose outstanding efforts have made this experiment 
possible. We thank the engineers and technicians for their work in constructing and maintaining 
the H1 detector, our funding agencies for financial support, the DESY technical staff for continual 
assistance and the DESY directorate for support and for the hospitality which they extend to the 
non DESY members of the collaboration.  We are also grateful to R.~Enberg, J.~R.~Forshaw, L.~Motyka 
and G.~Poludniowski for providing the results of the models and for useful discussions.



\begin{thebibliography}{99}

\bibitem{forshawryskin}
  J.~R.~Forshaw and M.~G.~Ryskin,
  Z.\ Phys.\ C {\bf 68} (1995) 137
  [hep-ph/9501376].
%
\bibitem{bartels}
  J.~Bartels, J.~R.~Forshaw, H.~Lotter and M.~W\"usthoff,
  Phys.\ Lett.\ B {\bf 375} (1996) 301
  [hep-ph/9601201].
%
\bibitem{zeushight}
  S.~Chekanov {\it et al.}  [ZEUS Collaboration],
  Eur.\ Phys.\ J.\ C {\bf 26} (2003) 389
  [hep-ex/0205081].
%
\bibitem{hightjpsi}
  A.~Aktas {\it et al.}  [H1 Collaboration],
  Phys.\ Lett.\ B {\bf 568} (2003) 205
  [hep-ex/0306013].
%
\bibitem{hightpsi}
  C.~Adloff {\it et al.}  [H1 Collaboration],
  Phys.\ Lett.\ B {\bf 541} (2002) 251
  [hep-ex/0205107].
%

\bibitem{electrosdme}
  C.~Adloff {\it et al.}  [H1 Collaboration],
  Phys.\ Lett.\ B {\bf 539} (2002) 25
  [hep-ex/0203022].
%
%

\bibitem{Serbo}
  I.~F.~Ginzburg, S.~L.~Panfil and V.~G.~Serbo,
  Nucl.\ Phys.\ B {\bf 284} (1987) 685.
%
\bibitem{ivanov}
  D.~Y.~Ivanov, R.~Kirschner, A.~Sch\"afer and L.~Szymanowski,
  Phys.\ Lett.\ B {\bf 478} (2000) 101
  [Erratum-ibid.\ B {\bf 498} (2001) 295]
  [hep-ph/0001255].
%
%
%
\bibitem{jeff1}
  R.~Enberg, J.~R.~Forshaw, L.~Motyka and G.~Poludniowski,
  JHEP {\bf 0309} (2003) 8
  [hep-ph/0306232].
%
\bibitem{jeff2}
  G.~G.~Poludniowski, R.~Enberg, J.~R.~Forshaw and L.~Motyka,
  JHEP {\bf 0312} (2003) 2
  [hep-ph/0311017].
%
\bibitem{Ball:1998ff}
  P.~Ball and V.~M.~Braun,
  Nucl.\ Phys.\ B {\bf 543} (1999) 201
  [hep-ph/9810475].
%
%
\bibitem{jeffnonrel}
  J.~R.~Forshaw and G.~Poludniowski,
  Eur.\ Phys.\ J.\ C {\bf 26} (2003) 411
  [hep-ph/0107068].
%
\bibitem{cox}
  B.~Cox, J.~R.~Forshaw and L.~L\"onnblad,
  JHEP {\bf 9910} (1999) 023
  [hep-ph/9908464].
%
\bibitem{enberg}
  R.~Enberg, private communication.
%
%
\bibitem{nim}
  I.~Abt {\it et al.}  [H1 Collaboration],
  Nucl. \ Instr. \ Meth. \ A {\bf 386} (1997) 310 and 348.
%
\bibitem{carl}
  C.~B.~Gwilliam, PhD Thesis, The University of Manchester, 2006,\\
  (available at http://www-h1.desy.de/publications/theses\_list.html)
%
\bibitem{jb}
  F. Jacquet and A. Blondel,
  DESY 79-048 (1979) 377.
%
\bibitem{angles}
  P.~Joos {\it et al.},
  Nucl.\ Phys.\ B {\bf 113} (1976) 53.
%
\bibitem{sdme}
  K.~Schilling and G.~Wolf,
  Nucl. \ Phys. \ B {\bf 61} (1973) 381.
%
\bibitem{diffvm}
  B.~List and A.~Mastroberardino, Proc. of the Workshop on Monte Carlo Generators for HERA Physics,
  eds.\, A.\,T.\,Doyle {\it et al.}, 
  DESY-PROC-1999-02 (1999) 396.
%
\bibitem{diffvmparam}
  C.~Adloff {\it et al.}  [H1 Collaboration],
  Z.\ Phys.\ C {\bf 75} (1997) 607
  [hep-ex/9705014].
%
\bibitem{goulianos}
  K.~Goulianos,
  Phys.\ Rept.\  {\bf 101} (1983) 169.
%
\bibitem{pdg}
  S.~Eidelman {\it et al.}  [Particle Data Group],
  Phys.\ Lett.\ B {\bf 592} (2004) 1.
%
\bibitem{soding}
  P.~S\"oding,
  Phys.\ Lett.\  {\bf 19} (1966) 702.
%
\bibitem{rs}
  M.~Ross and L.~Stodolsky,
  Phys. \ Rev. \ {\bf 149} (1966) 1173.
%
\bibitem{geant}
  R.~Brun {\em et al.},
  CERN-DD/DD-84-1.
%
%
\bibitem{omega}
  J.~Breitweg {\it et al.}  [ZEUS Collaboration],
  Phys.\ Lett.\ B {\bf 487} (2000) 273
  [hep-ex/0006013].
%
\bibitem{prime1}
  T.~J.~Killian {\it et al.},
  Phys.\ Rev.\ D {\bf 21} (1980) 3005.
%
\bibitem{prime2}
  W.~D.~Shambroom {\it et al.},
  Phys.\ Rev.\ D {\bf 26} (1982) 1.
%
%
%

\end{thebibliography}
\end{document}